\documentclass[runningheads]{llncs}
\usepackage{graphicx}
\usepackage{fixltx2e}
\usepackage{siunitx}
\usepackage{scrextend}
\usepackage{multirow}
\usepackage{subfigure}
\usepackage{floatrow}
\usepackage{wrapfig}
\usepackage{lipsum}

\begin{document}
\title{Cross-modality Knowledge Transfer for Prostate Segmentation from CT Scans}
\titlerunning{Automatic Prostate CT Segmentation using Synthetic CT}
\author{Yucheng Liu\inst{1} \and
Naji Khosravan\inst{2}\and
Yulin Liu\inst{1,3}\and
Joseph Stember\inst{1}\and
Jonathan Shoag\inst{4}\and
Christopher E. Barbieri\inst{4}\and
Ulas Bagci\inst{2} \and
Sachin Jambawalikar\inst{1}}
%index{Liu, Yucheng}
%index{Khosravan, Naji}
%index{Liu, Yulin}
%index{Stember, Joseph}
%index{Shoag, Jonathan}
%index{Bagci, Ulas}
%index{Jambawalikar, Sachin}
%\author{***}
%
\authorrunning{Liu et al.}
%\authorrunning{***}
\institute{***}
\institute{Department of Radiology, Columbia University Irving Medical Center, NY, NY,USA
\and
Center for Research in Computer Vision, University of Central Florida, FL, USA
\and
Department of Information and Computer Engineering, Chung Yuan Christian University, Taoyuan City, Taiwan
 \and
Department of Urology, New York Presbyterian Hospital, Weill Cornell Medical College, New York, NY, USA.}
\maketitle              % typeset the header of the contribution
\begin{abstract}
Creating large scale high-quality annotations is a known challenge in medical imaging. In this work, based on the CycleGAN algorithm, we propose leveraging annotations from one modality to be useful in other modalities. More specifically, the proposed algorithm creates highly realistic synthetic CT images (SynCT) from prostate MR images using unpaired data sets. By using SynCT images (without segmentation labels) and MR images (with segmentation labels  available), we have trained a deep segmentation network for precise delineation of prostate from real CT scans. For the generator in our CycleGAN,  the cycle consistency term  is used to guarantee that SynCT shares the identical manually-drawn, high-quality masks originally delineated on MR images. Further, we introduce a cost function based on structural similarity index (SSIM) to  improve the anatomical similarity between real and synthetic images. For segmentation followed by the SynCT generation from CycleGAN, automatic delineation is achieved through a 2.5D Residual U-Net. Quantitative evaluation demonstrates comparable segmentation results between our SynCT and radiologist drawn masks for real CT images, solving an important problem in medical image segmentation field when ground truth annotations are not available for the modality of interest.

\keywords{Domain adaptation  \and Deep learning \and CT synthesis \and prostate segmentation \and 2.5D \and Generative Adversarial Networks .}
\end{abstract}
\section{Introduction}
Prostate segmentation from radiology scans is often necessary for radiotherapy, prostatectomy, and calculation of prostate-specific antigen (PSA) density \cite{ref_lncs1}. Among imaging modalities, magnetic resonance imaging (MRI) provides the best soft tissue contrast and  yields the most accurate estimation on prostate volume, consistent with prostatectomy specimen volumes~\cite{ref_lncs2}. Unlike MRI, computed tomographic (CT) scans have difficulties to distinguish the boundaries of prostates and other adjacent tissues during segmentation~\cite{ref_lncs4}. Despite this, in current clinical practice, prostate radiation therapy dose calculations is primarily based on CT scans as it is the only modality that can derive electron density needed for the dosimetry calculations~\cite{ref_lncs6}. Therefore, planning systems generally require anatomical information to be delineated on CT scans.

% \begin{wrapfigure}[13]{R}{0.37\textwidth}
%\begin{center}
%\includegraphics[scale=0.50]{halfMRCT.png}
 %%\vspace{-0.5cm}
%\caption{An example slice of synthetic CT image (left) and its MR reference (right).\label{fig1}}
%\end{center}
%\end{wrapfigure}

In this study, we address a practical yet still very challenging issue of prostate segmentation from CT images when there are no ground truth CT annotations to supervise the segmentation algorithm. Instead, we target utilizing segmentation labels from widely available MRI data sets, and propose a two step  knowledge transfer algorithm to map the segmentation labels from MRI to CT scans. The correspondence between MRI to CT is established through a CycleGAN algorithm \cite{ref_lncs7} with a structural similarity preserving cost function. Highly realistic synthetic CT scans generated in the first step are then used to supervise a deep segmentation network in the second step. The training for the segmentation network is performed only on the synthetic images while testing is done on both synthetic and real CT scans for evaluation. While our framework does not enforce the use of any specific segmentation network to finalize the delineation process, we choose 2.5D Res-U-Net to accomplish this task with faster convergence, and higher accuracy.

%The synthetic CT image demonstrate rigid structure correspondence in bone and soft tissues, including fat, prostate, and rectum.

%In order to generate the optimized synthetic CT images, we changed the loss function of CycleGAN from Mean-square-error (MSE) to Structural similarity (SSIM) and fine-tuned the way of data augmentation. In this paper, we trained the segmentation network using SynCT data generated from different hyper-parameters and then evaluate the segmentation results on real CT images. 

\section{Methods}
\begin{figure}
\centering
\includegraphics[width=100 mm,scale=0.2]{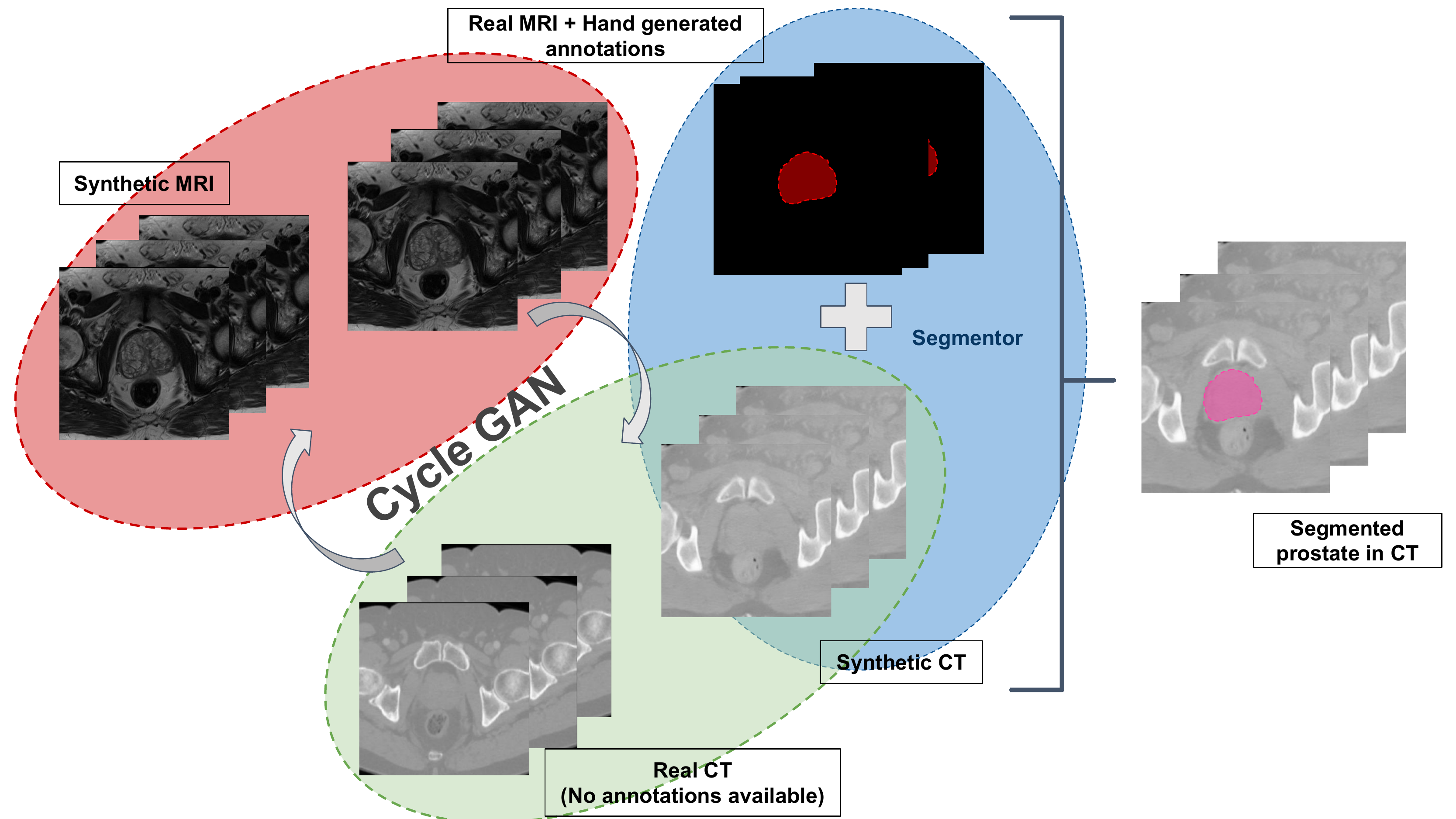}
\caption{\textbf{Workflow of CT image synthesis and automatic segmentation.}
The red box indicate the first step, CT image synthesis via CycleGAN model. SynCTs with identical anatomical structures as MRI were generated thus shared high-quality segmentation with MRI ( labeled red). The blue box indicate the second step, automatic segmentation via 2.5D Res-U-Net train with SynCT. The automatic generated segmentation (labeled pink) on true CT images were compared against manual segmentation from radiologist.  } 
\label{fig2}
\end{figure}

The proposed workflow includes two main steps as demonstrated in Figure~\ref{fig2}. First step is to generate high-quality and reliable CT images (SynCT) from MR images. Previous work \cite{ref_lncs9} has shown that domain adaptation from MR images to CT images is feasible using the CycleGAN architecture. We used a similar CycleGAN approach as baseline to create high-quality knowledge transfer between unpaired MRI and CT. %Also, they reported a lower Mean absolute error (MAE), which means more accurate synCT images results using unpaired data to train the network compared with paired data. 

Second step is to conduct automatic segmentation of prostate. We trained a U-Net based segmentation network to delineate the whole prostate area but with two main differences from the existing literature: (i) we used SynCT in training and real CT scans in testing, and (ii) we modified the U-Net~\cite{ref_lncs10} to increase the segmentation performance by adding residual blocks into the segmentation network. For better 3D information fusion, we also modified the segmentation architecture to utilize two additional adjacent slices in its input (i.e., 3-channel input).

\subsection{Data}
We used a total of three different data sets for our experiments and evaluations. For cycleGAN  training, $346$  T2 weighted MRI scans  from publicly available PROSTATEx-challenge data~\cite{ref_data1} was used. T2-weighted images were acquired using a turbo spin echo sequence with in-plane resolution of $0.4$-$0.6$ mm, slice thickness of $3.6$ mm and zero gap. Secondly, the testing data set for CycleGAN included 60 prostate MRI cases along with their high-quality delineation obtained from publicly available NCI-ISBI $2013$ challenge data \cite{ref_data2}. This data was used for generating the synthetic CT scans. We used 6-fold stratified cross validation for evaluation of the algorithms. Third, for real CT scans, as part of retrospective IRB approved study, we acquired prostate CT data from 120 anonymized patients  from our institution with resolution ($0.8\times0.8\times1mm^3$). CT intensity was clipped to $-500$HU to $500$HU to reveal more soft tissue contrast similar to a soft tissue CT window. Prostate MRI and CT data are completely different from each other, namely unpaired. Among in-house collected CT data, we chose $19$ of them to be manually segmented by a board certified radiologist for Dice score (DSC) comparison with our automatic segmentation method.
%\cite{ref_lncs13}. 

%30 patients were scanned by the PROSTATE-DIAGNOSIS collection (1.5T, resolution \(0.4\times0.4\times3 mm^3\) and the other 30 patients were scanned by the Prostate-3T collection (3T, resolution \(0.6\times0.6\times4 mm^3\). Ground-truth segmentation masks were provided by expert observers. 

\subsection{Synthetic CT Network: CycleGAN}
 The synthetic CT images were generated by the CycleGAN model~\cite{ref_lncs7}, which consisted of two pairs of generative adversarial networks (GAN) and two extra generators that convert generated data back to the original domain enforcing cycle consistency. In our study, the forward-direction GAN has a generator, \(G_{CT}(MR)\), that generate synthetic CT as real as possible such that a discriminator, \(D_{CT}\) cannot distinguish it from the real CT. The discriminator is to ensure the likeness of generated data with original data, hence, the reliability of the generated data heavily depends on the performance of the discriminator, the discriminator loss is described by Eq. 1. 
%\begin{figure}
%\centering
%\includegraphics[width=100mm,scale=0.25]{cGAN.png}
%\caption{\textbf{Forward loop of CycleGAN model.}
%The red box indicate the traditional generative adversarial network (GAN ) with a generator \(G_{CT}(MR)\) to generate synthetic CT and a discriminator \(D_{CT}\) to distinguish the true CT and the synthetic one. With one extra generator \(G_{MR}(SynCT)\) to convent SynCT back to MR domain, CycleGAN can compare and minimize the loss between true MRI and the reconstructed MRI despite no paired data available.  } \label{fig3}
%\end{figure}

\begin{equation}
\mathcal{L}_{D(CT)}  = \frac{1}{m}\sum_{j=1}^{m} [D_{CT}(I_{CT}^{j}-1)]^{2}+\frac{1}{n}\sum_{i=1}^{n} [D_{CT}(G_{CT}(I_{MR}^{i}))]^{2}
\end{equation}

Where \( I_{CT}^{j}\) denotes the j-th true CT slice; \(I_{MR}^{i}\) represents the i-th MRI slice; \(G_{CT}(I_{MR}^{i})\) represents the generated image by generator  \(G_{CT}(MR)\) from \(I_{MR}^{i}\); \(D_{CT}\) represents the discriminator who is trying to differentiate the generated image from CT images, if the discriminator cannot distinguish the generated image, it is labeled 1, which means the discriminator recognized this generated image as true CT image, otherwise a 0 label is given. 

The generator \(G_{MR}(SynCT)\) is translating the SynCT back to its' original data domain (MR domain). By minimizing the difference between the reconstructed data and the original data (cycle-consistency loss), a powerful constraint has been enforced on the model to prevent generated data deviation from ground-truth. The cycle-consistency loss is express as Eq. 2 here.

%\begin{equation}
%\mathcal{L}_{cycle}  = \frac{1}{n}\sum_{i=1}^{n} [I_{MR}^{i}-G_{MR}(G_{CT}(I_{MR}^{i}))]^{2}
%\end{equation}
%\begin{equation}
%SynCT  = G_{CT}(I_{MR}^{i})
%\end{equation}
\begin{equation}
\mathcal{L}^{SSIM}(P)  = \frac{1}{N}\sum_{p=1}^{N} [1-SSIM(p)]
\end{equation}
\begin{equation}
SSIM(p)  = (\frac{2\mu_{x}\mu_{y}+C_{1}}{\mu_{x}^{2}+\mu_{y}^{2}+C_{1}})(\frac{2\sigma_{xy}+C_{2}}{\sigma_{x}^{2}+\sigma_{y}^{2}+C_{2}}),
\end{equation}
where $P$ is the image patch, $N$ is number of pixels in $P$, and $p$ is the index of pixel; SSIM, for a pixel $p$, is defined as in Eq.3. Where $\mu_{x}, \mu_{y}$ and $\sigma_{x}, \sigma_{y}$ denotes mean pixel intensity and the standard deviations of pixel intensity in a local image patch centering at either $x$ or $y$. Also, $C_{1}$ and $C_{1}$ are small constants being added for stability. The cycle loss compares the reconstructed MRI with the true MRI slices in a pixel by pixel manner. In our new formulation, instead of computing mean-square-error (MSE), we propose to use structural similarity index (SSIM) that takes into account the context of the images at a higher level than pixel-level MSE~\cite{ref_lncs15}.

\subsection{Segmentation Network: 2.5D Res-U-Net}
The U-Net architecture\cite{ref_lncs10} has long skip connections to preserve spatial information during down-sampling. Besides long skip connections, short skip connections were also added forming residual blocks to prevent vanishing gradient and increase the convergence speed, the U-Net with short skip connections is called Res-U-Net~\cite{ref_lncs11}. Also, the proposed 2.5D input technique loads multiple slices simultaneously, which includes one central slice and its adjacent slices in out-of-plane direction. The number of channels is determined as the sum of central slice and the adjacent slices ($channel\:No. = central\:slice + adjacent\:slices$). The number of adjacent slices is defined through a designated context number which can query adjacent slices in both positive and negative directions ($adjacent\:slices = 2 \times context\:No$). For instance, if the context number is set to be 1, the selected adjacent slices will include +1 and -1 slices adjacent to the central slices. The context number can be adjusted in order to optimized the segmentation results.

%\subsection{Evaluations}
 
%The Dice similarity coefficient (DSC) is used to evaluate the quality of the generated segmentation outputs. DSC can compare the overlap area and the difference area between the ground-truth segmentation and the generated segmentation, as Eq. 6 shows here.

%\begin{equation}
%DSC=\frac{2|X\cap Y|}{|X|+|Y|}
%\end{equation}
%Where X denotes the ground-truth segmentation; Y denotes the generated segmentation. 

%voxel-based volume estimation

\section{Results}
The CycleGAN model was trained using Adam optimizer for 200 epochs with initial learning rate 0.0002; the 2.5D Res-U-Net model was trained using Adam optimizer for 300 epochs and binary cross entropy loss function was used because there are only two classes, masks and non-masks. Training took about 24 hours for CycleGAN to generate SynCT and about 12 hours for 2.5D Res-U-Net on a DGX-station with 4x Tesla V100 GPUs each with 32GB RAM. The segmentation results are displayed in Figure~\ref{fig4}. For data augmentation, rotation, flipping, and random crops from ratio 1 (no crop) to 0.5 (half crop) of original images were performed during training. 
\vspace{-0.2cm}
\begin{table}

\caption{Segmentation results (DSC) of MRI, SynCT and CT testing dataset.}\label{tab1}
%\begin{tabular}{L||C|L}
\begin{tabular}{>{\centering\arraybackslash}m{4.7cm}|>{\centering\arraybackslash}m{3.7cm}|>{\centering\arraybackslash}m{3.7cm}}

\textbf{Training dataset} & \textbf{Testing dataset} & \textbf{Dice score (DSC)}\\
\hline
MRI & MRI & $0.90\pm0.05$\\
\hline
SynCT &SynCT& $0.83\pm0.13$ \\
 & CT& $0.45\pm0.29$ \\
 \hline
Soft-tissue SynCT &SynCT&  $0.82\pm0.12$\\
 &CT & $0.62\pm0.15$\\
 \hline
Soft-tissue SynCT \par Data augmentated &SynCT& $0.65\pm0.09$\\
 &CT& $0.68\pm0.09$\\
 \hline
Soft-tissue SynCT \par Data augmentated \par SSIM loss & SynCT & $0.80\pm0.12$\\
&CT& $0.73\pm0.09$\\
%\hline
\end{tabular}
\end{table}
\vspace{-0.5cm}

2.5D Res-U-Net trained and tested on MRI data illustrates the upper bounds of performance, network trained on CT/SynCT data will intuitively be lower than 0.9 (Table~\ref{tab1}). SynCTs paired with MRI segmentations were used to train the automatic segmentation network. For SynCT generated from default CycleGAN setting (MSE loss, random crop with fix ratio, 284 to 256 pixels) and no intensity clipping, we achieved 0.83\(\pm 0.13\) and 0.45\(\pm 0.29\) DSC for SynCT and CT testing set, respectively; for Soft-tissue SynCT (intensity clipped from -500 HU to 500 HU), we achieved 0.82\(\pm 0.12\) and 0.62\(\pm 0.15\) DSC for SynCT and CT testing set, respectively. More aggressive data augmentation (random crop with random ratio, rotation, flipping) also adapted to generate higher quality SynCT from CycleGAN, which achieved 0.65\(\pm 0.09\) and 0.68\(\pm 0.09\) DSC for SynCT and CT segmentation testing set, respectively. To increase the structure accuracy, the cycle loss has replaced into structural similarity index (SSIM), the 2.5D Res-U-Net trained with SynCT-SSIM achieved 0.80\(\pm 0.12\) and 0.73\(\pm 0.09\) DSC for SynCT and CT testing set, respectively. Note that the DSC of SynCT decrease and the DSC of CT increase to reach a compatible point with no statistical difference (\(p > 0.05\)), also the standard deviations are converging. This tendency indicated our SynCT gradually reached a point where there was no difference with true CT from 2.5D Res-U-Net network perspective.

\section{Discussion and Concluding Remarks}
\begin{figure}
%\centering
\includegraphics[width=90mm,scale=0.25]{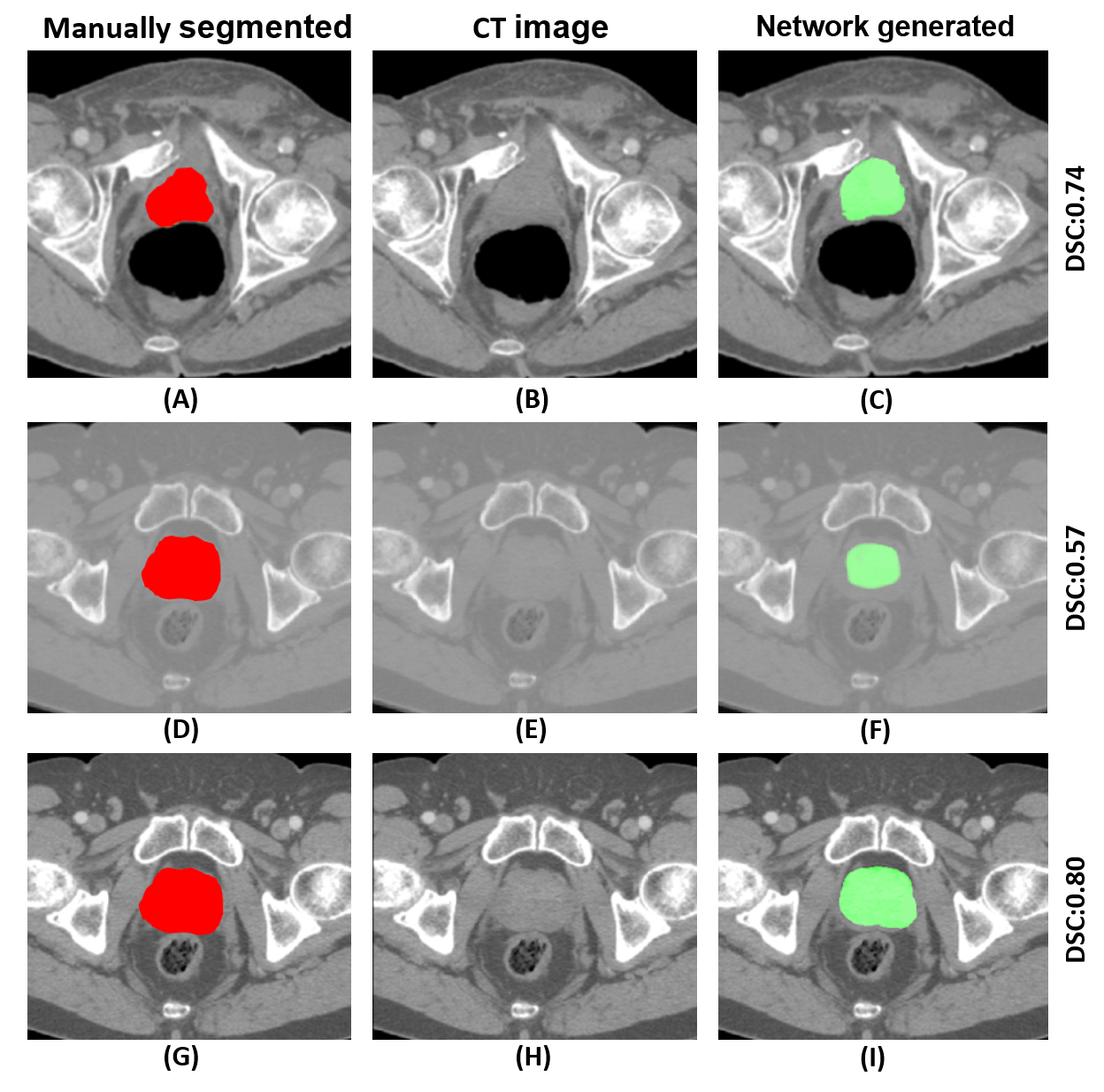}
\caption{Example slices of segmentation results on true CT.
(A) Under-segmented prostate by expert radiologists. 2.5D Res-U-Net can generate better segmentation (C) since it adapted the segmentation from MRI, however, resulting a misleadingly lower DSC, 0.74. CT with normal intensity can vary from -1000 HU (air) to 1000 (bone), therefore soft tissues consists of similar HU numbers may not be seen clearly on the images, as demonstrated on the middle part of the figure, where (D) is CT with ground-truth segmentation from radiologist, (E) is CT without any intensity adjustment, and (F) is CT with 2.5D Res-U-Net generated segmentation. Last row demonstrates CT with soft tissue window (-500 HU to 500 HU, we called ST-CT (soft tissue CT)), which is slightly larger than typical soft  tissue window, -150 HU to 350 HU, to accommodate more information in the slices. Where (G) is  ST-CT with ground-truth segmentation, (H) is the ST-CT, and (I) is the ST-CT with 2.5D Res-U-Net generated segmentation. At the same case, the DSC of CT and ST-CT is 0.57 and 0.80, respectively.} \label{fig4}
\end{figure}

Intensive studies have been made regarding prostate CT automatic segmentation. Recently, the reported highest DSC is \(0.88 \pm 0.03\) by Liu \textit{et al.}~\cite{ref_lncs17}using U-Net and 1114 ture CT cases. Our average result is \(0.73 \pm 0.09\) which is compatibe with Burgos \textit{et al.}~\cite{ref_lncs18} using multi-atlas based SynCT (0.73 DSC). We have shown that the SynCT and the CT testing results have no statistical difference indicating the feasibility of using SynCT to train a neural network for a very challenging segmentation task. In some cases DCS is low but not due to low performance of the proposed network. The low DSC is sometimes due to noise in the contouring in the hand-drawn CT ground-truth segmentation and large anatomical and pathological variations (see Figure~\ref{fig4}).

%\vspace{-.9cm}

\textbf{Data Augmentation:} We used MRI and CT scans from different data sources, MRI have smaller field-of-view (FOV) compared to CT.  Inconsistent FOV encouraged CycleGAN to shift the anatomy without focusing on anatomical details.  To generate high-quality SynCT, we central cropped the CT images by 50\%  to remove the surrounding air and scanning table. Then augment the data with random ratio (1 - 0.5) random crop, rotation, and flipping to reduce certain geometry tendency affecting the learning process. 

\begin{figure}
\centering
\includegraphics[width=120mm,scale=1.5]{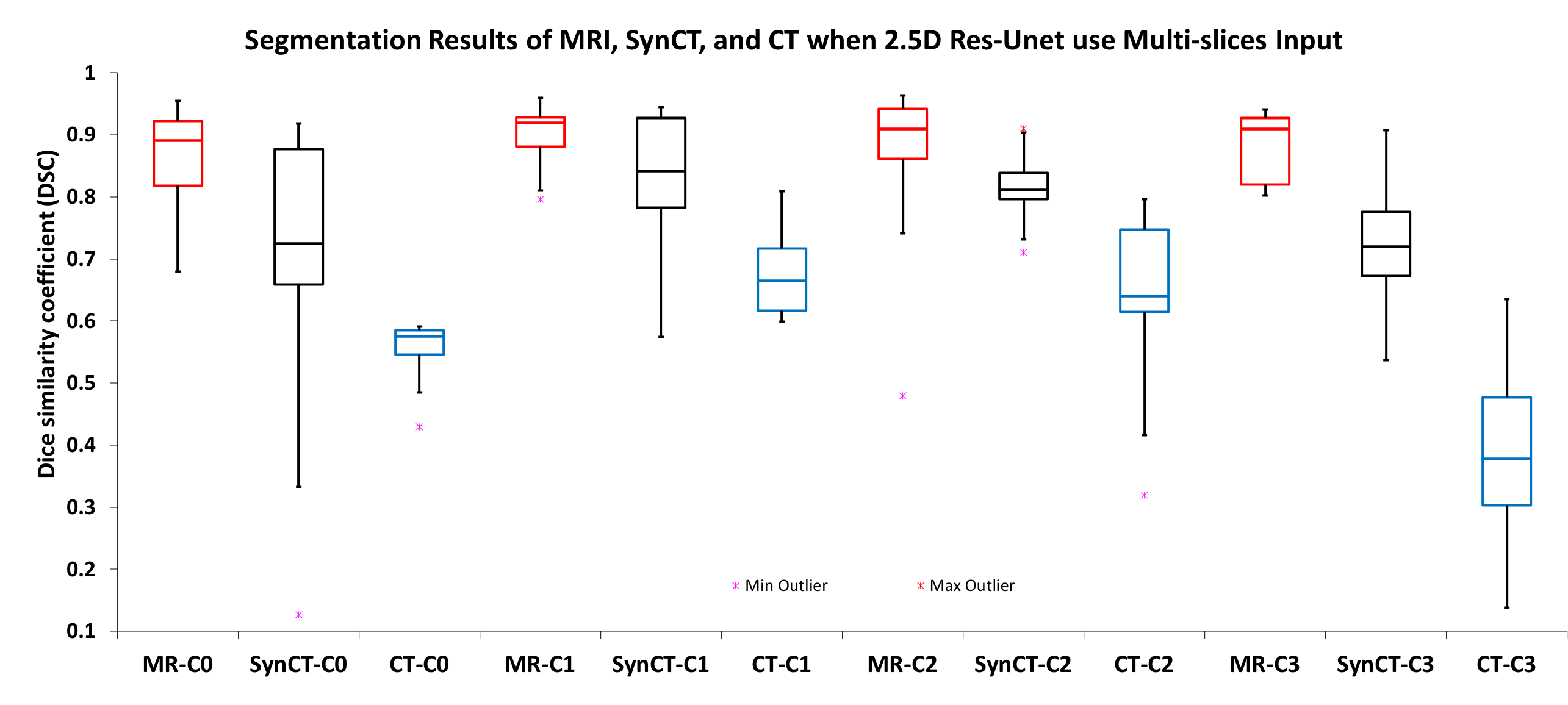}
\caption{Boxplots are showing the Dice scores for prostate segmentation from MRI, SynCT, and CT, respectively.} \label{fig3}
\end{figure}

\textbf{2.5D Technique:} 2.5D multi-slices input technique can affect the segmentation network performance as Figure~\ref{fig3} shows here. For SynCT, from single slice to 3-slices, DSC increases significantly (\(p<0.05\)) by \(19.11\%\), from 3-slices to 5-slices no significant difference was found, from 5-slices to 7-slices, DSC decreased \(12.5\%\); for CT, from single slice to 3-slices, DSC increase significantly by \(24.17\%\), from 3-slices to 5-slices no significant difference found, from 5-slices to 7-slices, DSC drop significantly by \(40.93\%\). Therefore, to optimized the performance of 2.5D Res-U-Net and also save training time, context number 1 (3-slices input) was used for all experiments.

%CT-C0 (single slice) and CT-C1 (3 slices); from  CT-C1 (3 slices) and CT-C3 (7 slices). One-tailed T-test shows significant difference between SynCT-C0 and SynCT-C1, CT-C0 and CT-C1, CT-C1 and CT-C3. 

In summary, we proposed a novel approach to segment prostate from CT scans when the  ground-truth was absent. Synthetic CT scans that share high-quality segmentation with MRI were used to train a deep-learning based automatic segmentation network (2.5D Res-U-Net). The testing results on true CT achieved 0.73 DSC which is comparable with SynCT.  We also examined and identified the optimal numbers of multiple slices input, which are 3 or 5 slices. Future steps will include 3D volume assessment and continue improvement of the quality of synthetic CT generation.

%\bibliographystyle{splncs04}
%\bibliography{mybibliography}
%

%evaluations: 
%https://www.researchgate.net/publication/259202304_New_technique_for_prostate_volume_assessment

\end{document}